\documentclass[twocolumn,pra,aps,showpacs,nofootinbib]{revtex4}
\usepackage{graphicx}

\begin{document}


\preprint{LA-UR-01-1909}
\title{How do two observers pool their knowledge about a quantum system?%
      }

\author{Kurt Jacobs}

\affiliation{T-8, Theoretical Division, MS
B285, Los Alamos National Laboratory, Los Alamos, New Mexico 87545}

\begin{abstract}
In the theory of classical statistical inference one can derive a simple rule by which two or more observers may combine {\em independently} obtained states of knowledge together to form a new state of knowledge, which is the state which would be possessed by someone having the combined information of both observers. Moreover, this combined state of knowledge can be found without reference to the manner in which the respective observers obtained their information. However, we show that in general this is not possible for quantum states of knowledge; in order to combine two quantum states of knowledge to obtain the state resulting from the combined information of both observers, these observers must also possess information about how their respective states of knowledge were obtained. Nevertheless, we emphasize this does not preclude the possibility that a unique, well motivated rule for combining quantum states of knowledge without reference to a measurement history could be found. We examine both the direct quantum analogue of the classical problem, and that of quantum state-estimation, which corresponds to a variant in which the observers share a specific kind of prior information. 
\end{abstract}


\maketitle

\section{Introduction}
Consider a variable $N$, which can take a range of values, but which in reality has a single value. An observer's {\em state of knowledge} regarding the value of $N$ may be written as a probability density which is a function of the possible values~\cite{Jaynes}. Specifically, if $N$ can take integer values in a range $n = 1,\ldots , n_{\mbox{\scriptsize max}}$, then we write the probability density as $P(n)$ (where naturally $\sum_n P(n) = 1$ and $P(n) \ge 0, \;\forall n$). We will refer to $P(n)$ as being a {\em classical} state of knowledge.

An observer's state of knowledge regarding $N$ may be thought of as a result of that observer having obtained some information about the value of $N$, and the more information she obtains, the lower is the entropy, $S(P(n))$ of her density. If we now consider two observers, each of whom have obtained independent information about the value of $N$, then in general they will have different states of knowledge. Now, since we have made the assumption that their information is independent, if they pool that knowledge, then they should end up with a combined state of knowledge which has an entropy that is lower than either of their two individual states of knowledge. Using Bayesian statistical inference, one can obtain the rule by which the two observers can combine their respective probability densities to obtain the new probability density which describes what they jointly know about $N$. That is, under the assumption that the information of the respective observers is independent, their joint state of knowledge may be obtained entirely from their individual states of knowledge, whith no reference to how the knowledge was obtained. However, when one is concerned with quantum states of knowledge, which are given by density matrices rather than probability densities, it turns out that the situation is quite different. 

In the following section we describe the rule for combining classical states of knowledge. In section III we consider how the situation is different for quantum states of knowledge, examining both the direct quantum analogue of the classical problem, and that of quantum state-estimation. In section IV we conclude with a brief summary.

\section{Combining classical states of knowledge}
Bayes' theorem provides a means to describe how information is obtained about a random variable $N$~\cite{Bayes}. An observer makes a measurement and obtains a result $M$, which is related to the variable $N$ through the conditional probability $P(m|n)$. If the observer's state of knowledge regarding $N$ before the measurement was $P(n)$, then after the measurement it is 
\begin{equation}
  P(n|m) = \frac{P(m|n)P(n)}{P(m)}
\end{equation} 
where $P(m)$ may be determined completely from the fact that $P(n|m)$ must be normalized. If $K$ measurements are made, with independent results $m_k$, then the observer's final state of knowledge is given by 
\begin{equation}
  P(n|m_1,\ldots ,m_K) = \frac{1}{{\cal N}(\{m_k\})}\left[ \prod_{k=1}^{K} P(m_k|n) \right] P(n) 
\label{Bform}
\end{equation}
where ${\cal N}(\{m_k\})$ is the overall normalization, given by ${\cal N}(\{m_k\}) = \prod_{k=1}^{K}P(m_k)$. 

Consider two observers, Alice and Bob, where Bob's state of knowledge, $P(n|m_1,\ldots ,m_K)$ is the result of obtaining measurement results $\{m_k\}$, and Alice's, $Q(n|l_1,\ldots ,l_J)$, is the result of obtaining measurement results $\{l_j\}$. By asserting that their respective bodies of information about $N$ are given {\em entirely} by the measurement results, we must take the initial state of knowledge in the formula given by Eq.(\ref{Bform}) to be given by a flat distribution, $P(n) = 1/n_{\mbox{\scriptsize max}}$. This is because, by the assertion, all information (that is, everything that contributes to making each observer's state of knowledge possess less than maximum entropy) is embodied in the measurement results. Therefore, the initial state of knowledge must be the one with maximum entropy, which is $P(n)=1/n_{\mbox{\scriptsize max}}$.

The state of knowledge of a third observer, Charlie,  who has the combined information of both Alice and Bob, is given by
\begin{widetext}
\begin{eqnarray}
  P(n|\{m_k\},\{l_j\}) & = & \frac{1}{{\cal N}(\{m_k\},\{l_j\})}\left[ \prod_{k=1}^{K} P(m_k|n) \right] \left[ \prod_{j=1}^{J} P(l_j|n) \right] (1/n_{\mbox{\scriptsize max}}) \nonumber \\
                & = &  f(\{m_k\},\{l_j\}) P(n|m_1,\ldots ,m_K)Q(n|l_1,\ldots ,l_J)
\end{eqnarray}
\end{widetext}
where $f$ is some function of the measurement results, and may be determined entirely from the requirement of normalization. 
Thus, to combine the states of knowledge of the two observers, we simply multiply the respective probability densities and renormalize.

It is possible to write Bayes theorem in the same form as the theory of generalized quantum measurements, since this generalized theory must reduce to Bayes theorem under certain conditions. If we write an initial state of knowledge as a diagonal matrix $\rho$, where the diagonal matrix elements $\rho_{n,n}$ are given by $P(n)$, and we write the $M$ conditional probabilities $P(m|n)$ as diagonal matrices $E_m$ where the diagonal elements are given by $(E_m)_{n,n} = P(m|n)$, then using Bayes theorem the final state of knowledge, after obtaining measurement result $m$ is given by
\begin{eqnarray}
  \rho_m = \frac{E_m \rho}{P(m)} = \frac{\sqrt{E_m} \rho \sqrt{E_m}}{\mbox{Tr}[E_m\rho]}
\end{eqnarray}
where $\sum_m E_m = I$. After the measurement, and conditional upon the measurement result, one can also perform, for example, a reversible deterministic transformation on the variable being estimated. The result of that is to multiply the state of knowledge on the left by a matrix $T_m$, and on the right by $T_m^{\mbox{\scriptsize T}}$, where $T_m$ has every element zero, except for exactly one element in each row and column, which is unity.

To obtain the theory of generalized quantum measurements one relaxes the restriction that the matrices $\rho$ and $E_m$ be diagonal (but retains the restriction that they be positive). After the measurement, and conditional upon the outcome, one is allowed to perform a unitary transformation, $U_m$, on the system, and this generalizes the deterministic classical transformation $T_m$ introduced above. This gives
\begin{eqnarray}
  \rho_m = \frac{U_m\sqrt{E_m} \rho \sqrt{E_m}U^\dagger_m}{\mbox{Tr}[E_m\rho]} .
\end{eqnarray} 
However, this final unitary tells us nothing about the information gathering process, just as the performance of the deterministic transformation on a classical system following a classical measurement tells us nothing about the classical information gathering process. Hence one can dispense with the unitaries if one is interested only in the process of obtaining information. In what follows we will refer to the operators $E_m$ as the {\em effects}~\cite{Kraus}, following Kraus.

With this notation the rule for combining classical states of knowledge is 
\begin{equation}
  \rho_{\mbox{\scriptsize tot}} = \frac{\rho_{\mbox{\scriptsize A}}\rho_{\mbox{\scriptsize B}}}{\mbox{Tr}[\rho_{\mbox{\scriptsize A}}\rho_{\mbox{\scriptsize B}}]}, \;\;\; ([\rho_{\mbox{\scriptsize A}},\rho_{\mbox{\scriptsize B}}] = 0) .
\end{equation}
As we will see in the next section, the situation for quantum states is significantly more complex, however. 

\section{Combining quantum states of knowledge}
\subsection{The direct quantum analogue}
In the case of classical measurements considered above, each observer performed measurements on the system in question. Since in that case all the measurement operators for the different observers commute, we did not have to worry about the order in which the measurements were performed. In the quantum case however, the states of knowledge of the two observers, after making their measurements, are dependent upon the order in which the measurements are made, and moreover, also due to the non-commutativity, the state of knowledge of one observer will in general depend upon the choice of measurements made by the other observer.
As an example, Alice and Bob's states of knowledge, after Alice has made $N$ measurements, given by POVM's $\sum_{i_1}  A^{1\dagger}_{i_1} A^{1}_{i_1} = 1 ,\ldots, \sum_{i_N} A^{N\dagger}_{i_N} A^{N}_{i_N} = 1$, and Bob has made $M$ measurements, given by POVM's $\sum_{i_1}  B^{1\dagger}_{j_1} B^{1}_{j_1} = 1 ,\ldots, \sum_{j_M} B^{M\dagger}_{j_M} B^{M}_{j_M} = 1$, are, respectively,
\begin{widetext} 
\begin{eqnarray}
  \rho_{\mbox{\scriptsize A}} & = & \frac{1}{{\cal N}_{\mbox{\scriptsize A}}} \sum_{j_1=1}^{j_1^{\mbox{\tiny max}}}\cdots\sum_{j_M=1}^{j_M^{\mbox{\tiny max}}} {\cal P} (A^{1}_{i_1},\ldots, A^{N}_{i_N},B^{1}_{j_1},\ldots,B^{M}_{j_M}) \left[ {\cal P} (A^{1}_{i_1},\ldots, A^{N}_{i_N},B^{1}_{j_1},\ldots,B^{M}_{j_M}) \right]^\dagger \label{fullrho1} \\
  \rho_{\mbox{\scriptsize B}} & = & \frac{1}{{\cal N}_{\mbox{\scriptsize B}}} \sum_{i_1=1}^{i_1^{\mbox{\tiny max}}}\cdots\sum_{i_N=1}^{i_N^{\mbox{\tiny max}}} {\cal P}(A^{1}_{i_1},\ldots, A^{N}_{i_N},B^{1}_{j_1},\ldots,B^{M}_{j_M}) \left[ {\cal P}(A^{1}_{i_1},\ldots, A^{N}_{i_N},B^{1}_{j_1},\ldots,B^{M}_{j_M}) \right]^\dagger \label{fullrho2}  
\end{eqnarray}
\end{widetext}
where ${\cal P}$ denotes a given permutation of the product of its arguments, the permutation being determined by the order in which the measurements were made. Alice's state of knowledge is given by averaging over Bob's measurement results and vice versa. Note that we can always rewrite the history of measurements by Alice and Bob as a single POVM in which the elements have two indices. For example: 
\begin{eqnarray}
  \rho_{\mbox{\scriptsize A}} & = & \frac{1}{{\cal N}_{\mbox{\scriptsize A}}} \sum_{j=1}^{j^{\mbox{\tiny max}}} {\cal A}_{ij}{\cal A}^\dagger_{ij} \\
  \rho_{\mbox{\scriptsize B}} & = & \frac{1}{{\cal N}_{\mbox{\scriptsize B}}} \sum_{i=1}^{i^{\mbox{\tiny max}}}{\cal A}_{ij}{\cal A}^\dagger_{ij}
\end{eqnarray}
where
\begin{eqnarray}
    i^{\mbox{\tiny max}} = \prod_{k=1}^{N} i_k^{\mbox{\tiny max}} , \;\;\;\;
    j^{\mbox{\tiny max}} & = & \prod_{k=1}^{M} j_k^{\mbox{\tiny max}} .
\end{eqnarray}
We could also include in the expressions for $\rho_{\mbox{\scriptsize A}}$ and $\rho_{\mbox{\scriptsize B}}$ measurements made by another observer, Eve, the results of which neither Alice nor Bob have access too. This involves simply choosing one or more of the indices and allowing both Alice and Bob to sum over them. While in the classical case such an addition makes no difference to the observers' final states of knowledge, quantum mechanically it does. We will include measurements by Eve in what follows when relevant to the discussion.

Before considering the problem of combining states of knowledge, it is worth discussing when two states of knowledge are {\em consistent} with one another. If we take as our basic assumption that two separate observers obtain their respective states of knowledge in the above manner (that is, by each making measurements at various times, and averaging over the results of the other observer's measurements), then two states of knowledge are {\em consistent} with 
one another if and only if they may be written in the form given by Eqs.(\ref{fullrho1}) and (\ref{fullrho2}) (with the addition of measurements by Eve over which both observers sum)~\footnote{Of course, in considering the consistency of states of knowledge, it is essential to include the case in which the observers' states are {\em not} independently obtained. This can be achieved by providing indices in Eqs. (\ref{fullrho1}) and (\ref{fullrho2}) which are summed over by {\em neither} observer. However, this addition does not change the remainder of the argument.}. By an inspection of the form of these equations, it is immediately clear that the two densities matrices are given each by a sum of terms, where the two sums have at least one term in common (exactly one term if Eve makes no measurements, more than one if she does). Thus we can always write 
\begin{eqnarray}
  \rho_{\mbox{\scriptsize A}} & = & \alpha\sigma + \sum_{k=1}^{K} p^{\mbox{\scriptsize A}}_k |\phi^{\mbox{\scriptsize A}}_k \rangle\langle\phi^{\mbox{\scriptsize A}}_k| \label{common1} \\
  \rho_{\mbox{\scriptsize B}} & = & \beta\sigma + \sum_{l=1}^{L} p^{\mbox{\scriptsize B}}_l |\phi^{\mbox{\scriptsize B}}_l \rangle\langle\phi^{\mbox{\scriptsize B}}_l| 
  \label{common2}
\end{eqnarray}
Note that in these equations, normalization of the density matrices implies $\alpha = 1 - \sum_k p^{\mbox{\scriptsize A}}_k$ and $\beta = 1 - \sum_l p^{\mbox{\scriptsize B}}_l$. Now, Brun {\em et al.}~\cite{Brun} have also shown that the reverse is true. That is, that for two density matrices that can be written as a sum of terms with at least one term in common, a set of POVM's may be constructed such that two observers can obtain those density matrices by making measurements. We will refer to this as the realizability property of quantum measurements. (Actually, Brun {\em et al.} only show this when $\sigma$ is pure - however, extending their method to mixed states is straightforward, and this is included in our analysis below.) Hence, we may conclude that two density matrices are consistent if and only if an expansion may be found for each such that these expansions have a term in common. Thus, by using a slightly different starting point (Eqs.(\ref{fullrho1}) and (\ref{fullrho2})) than that used in reference~\cite{Brun}, we arrive at the same condition for consistency as found there.

The problem of obtaining the state of knowledge of a third observer, Charlie, who has access to all the information (that is, both Alice and Bob's measurement results) is now the following: given the two states $\rho_{\mbox{\scriptsize A}}$ and $\rho_{\mbox{\scriptsize B}}$, we must find the state
\begin{eqnarray}
  \rho_{\mbox{\scriptsize C}} & = & \frac{1}{{\cal N}_{\mbox{\scriptsize C}}} {\cal A}_{ij}{\cal A}^\dagger_{ij}
\end{eqnarray}
since this is Charlie's state of knowledge. The problem with obtaining such a state is that, given merely $\rho_{\mbox{\scriptsize A}}$ and $\rho_{\mbox{\scriptsize B}}$, $\rho_{\mbox{\scriptsize C}}$ is not well-defined. That is, fixing $\rho_{\mbox{\scriptsize A}}$ and $\rho_{\mbox{\scriptsize B}}$, one can find two different sets of measurement histories that give two different values for $\rho_{\mbox{\scriptsize C}}$. We can show that this is the case by employing the method used by Brun {\em et al.}~\cite{Brun} to prove the realizability property mentioned above, along with another result also contained in~\cite{Brun}. A proof of the realizability property is as follows. 

We consider three systems, the system of interest to the two observers, $S$, and two auxiliary systems, $S_{\mbox{\scriptsize A}}$ and $S_{\mbox{\scriptsize B}}$. We choose the initial state of the three systems to be proportional to the identity, in keeping with the constraint that the observers share no prior information. Next, the two observers both make measurements which project the composite system into a pure state, which is (up to an overall normalization factor),
\begin{eqnarray}
 |\Psi\rangle & \propto & \sum_{i=1}^{N} \sqrt{\lambda_i} |\phi_i \rangle |{\mbox{A}}_i\rangle |{\mbox{B}}_i\rangle \nonumber \\ 
              & + & \sum_{k=1}^K \sqrt{p^{\mbox{\scriptsize A}}_k/\alpha} |\phi^{\mbox{\scriptsize A}}_k \rangle |{\mbox{B}}_{(k+N)}\rangle |\psi^{\mbox{\scriptsize A}} \rangle \nonumber \\ 
              & + & \sum_{l=1}^L \sqrt{p^{\mbox{\scriptsize B}}_l/\beta} |\phi^{\mbox{\scriptsize B}}_l \rangle |{\mbox{A}}_{(l+N)}\rangle |\psi^{\mbox{\scriptsize B}} \rangle .
\end{eqnarray}
The ${\lambda_i}$ and the $|\phi_i\rangle$ are respectively the eigenvalues and eigenvectors of $\sigma$:
\begin{equation}
 \sigma = \sum_{i=1}^{N} \lambda_i |\phi_i\rangle \langle\phi_i| .
\end{equation}
The sets of states $\{|{\mbox{A}}_i\rangle\}$ and $\{|{\mbox{B}}_i\rangle\}$ are orthonormal bases for the systems $S_{\mbox{\scriptsize A}}$ and $S_{\mbox{\scriptsize B}}$, respectively, and
\begin{eqnarray}
  |\psi^{\mbox{\scriptsize A}} \rangle & = & \frac{1}{\sqrt{N}} \sum_{i=1}^N  |{\mbox{A}}_i \rangle , \\
  |\psi^{\mbox{\scriptsize B}} \rangle & = & \frac{1}{\sqrt{N}} \sum_{i=1}^N  |{\mbox{B}}_i \rangle . 
\end{eqnarray}
Alice makes now a measurement projecting $S_{\mbox{\scriptsize A}}$ onto the basis $\{|{\mbox{A}}_i\rangle\}$, obtaining (with a probability less than unity) a result associated with the one of the states $|{\mbox{A}}_n\rangle$, where $n \in 1, \ldots, N$. The probability that she obtains result $n$ is 
\begin{equation} 
  P(n) = \frac{1}{\langle \Psi | \Psi \rangle} \left( \lambda_n + \frac{1-\alpha}{\alpha N} \right).
\end{equation}
In addition, having no access to system $S_{\mbox{\scriptsize B}}$, she traces over it. After obtaining result $n$, and tracing over system $S_{\mbox{\scriptsize B}}$, her state of knowledge regarding system $S$ is (up to a normalization factor) 
\begin{equation}
 \rho^{\mbox{\scriptsize A}}_n \propto \lambda_n |\phi_n\rangle \langle\phi_n| + \frac{1}{\alpha N} \sum_{k=1}^K p^{\mbox{\scriptsize A}}_k |\phi^{\mbox{\scriptsize A}}_k \rangle \langle \phi^{\mbox{\scriptsize A}}_k | .
\end{equation}
At this point she throws away the information about which of the $N$ results she obtained, so her state is the result of averaging the $\rho^{\mbox{\scriptsize A}}_n$ over $n$. Her final state of knowledge regarding system $S$ is then precisely that given by Eq.(\ref{common1}). Bob performs an equivalent procedure, but this time measuring system $S_{\mbox{\scriptsize B}}$, projecting it onto the basis $\{|{\mbox{B}}_i\rangle\}$, and throwing away the information about which of the results $i = 1, \ldots , N$ was obtained. His final state of knowledge regarding $S$ is that given by Eq.(\ref{common2}). While Alice's and Bobs states are important, just as important for our purposes is Charlie's state of knowledge (where, as above, Charlie is an observer with access to the measurement results of both Alice and Bob)). For the above measurement procedure, Charlie's state of knowledge is $\sigma$. 

To complete our analysis, we need a second result shown in~\cite{Brun}, which is the following: Two density matrices may be written in the form given by Eqs. (\ref{common1}) and (\ref{common2}), with $\sigma = |\phi\rangle\langle\phi|$, if and only if the state $|\phi\rangle$ is contained within the space which is the intersection of the supports of $\rho_{\mbox{\scriptsize A}}$ and $\rho_{\mbox{\scriptsize B}}$ (It is also a simple matter to extend this result to the case when $\sigma$ is mixed - in this case, it is the {\em support} of $\sigma$ which must be contained within the intersection of $\rho_{\mbox{\scriptsize A}}$ and $\rho_{\mbox{\scriptsize B}}$). Note that the support of a density matrix is the space spanned by its non-zero eigenvectors. 

Combining this with the previous result provides us with the following procedure. Given two states of knowledge, who's supports have an intersection which is at least two dimensional, we can can choose two distinct states from this intersection. For each of these distinct states we can construct a measurement procedure for Alice and Bob such that Charlie's resulting state of knowledge is the state in question. Charlie's state of knowledge is therefore dependent upon the measurement history and is not well defined by the states of knowledge of Alice and Bob alone. In particular we can state the following lemma.

\vspace{1mm}
{\bf Lemma}: Given two consistent, independently obtained states of knowledge, held respectively by observers A and B, the state of knowledge of a third observer C, who has access to the information of both A and B, may be any (possibly mixed) state, who's support lies within the intersection of the supports of the states of A and B.
\vspace{1mm}

Thus, the problem of combining density matrices, in the sense of obtaining the state resulting from the full combined information of both observers, is not well defined. However, we would like to finish this section by pointing out that it may well still be possible to obtain a well-defined procedure for forming a combined state of knowledge from two independently obtained density matrices. Consider the situation in which two observers posses states of knowledge, but wish to combine them without reference to how the states where obtained. The observers should then take into consideration all the states which could result from fully combining their information; that is, they should take into account all possible measurement histories consistent with their states of knowledge.  Given one of these possible measurement histories, call it $\Lambda$, there will be a certain probability that the observers' respective states of knowledge result from this. Specifically, if $\Lambda$ is the POVM described by the set of operators ${\cal A}_{ij}$, then the probability that the states of knowledge corresponding to Alice obtaining outcome $i$ and Bob obtaining outcome $j$ result from this measurement, is $P(\rho_{\mbox{\scriptsize A}},\rho_{\mbox{\scriptsize B}}|\Lambda) = \mbox{Tr}[{\cal A}_{ij}{\cal A}_{ij}^\dagger]$. The state of knowledge of an observer who knows both outcomes is $\sigma_\Lambda = {\cal A}_{ij}{\cal A}_{ij}^\dagger$, suitably normalized. The combined state, resulting from a knowledge only of $\rho_{\mbox{\scriptsize A}}$ and $\rho_{\mbox{\scriptsize B}}$, would then be (using Bayes' theorem)
\begin{eqnarray}
  \sigma_{\mbox{\scriptsize AB}} & = & \frac{1}{\cal N} \int_{\Lambda} P(\rho_{\mbox{\scriptsize A}},\rho_{\mbox{\scriptsize B}}|\Lambda) \; \sigma_\Lambda  \; d\Lambda \nonumber \\
   & = & \frac{1}{\cal N} \int_{\Lambda} {\cal A}_{ij}{\cal A}_{ij}^\dagger  \; d\Lambda
\end{eqnarray} 
The central point is that, while any state in the intersection of the supports of the two density matrices is a {\em possible} state for Charlie, not all these states will be equally {\em likely}. Of course, such a procedure would require a well motivated choice of measure over the measurement histories $\Lambda$.


\subsection{Quantum state-estimation}
When there exists more than one copy of a system prepared in a given (but unknown) quantum state, then measurements can be made on each of the systems in order to discover the nature of the state. This procedure is referred to as quantum state estimation. As will be made clear in what follows, this is a departure from the formulation considered in the previous section, because the initial state of the combined system (that is, the system consisting of all the identically prepared systems combined, before any measurements have been made), is no longer the identity. That is, the observers share some initial knowledge about the state of the combined system.

In quantum state estimation, a number of copies, $N$, of a given quantum state $|\psi\rangle$ are prepared, but the observer who wishes to perform the state estimation by making measurements on one or more of these copies has incomplete knowledge as to what the state is. Thus one may write the observer's initial state of knowledge as a probability density, $P(|\psi\rangle)$, over the possible states. As measurements are made on the various copies, the observer's state of knowledge changes, and eventually the probability density becomes sharply peaked about the true state~\cite{Jones} (see also~\cite{Massar,Hradil,Slater,Derka,Buzek,Tarrach,Banaszek,Gill}). In this case, the quantum state can now be viewed like a classical variable being estimated, and Bayes' rule can be applied to $P(|\psi\rangle)$ as each subsystem is measured. But in the previous section we asserted that an observer's state of knowledge is given by a density matrix, and now we are saying that it should be encoded as a probability density over pure states! Are we contradicting ourselves? Not at all. Recently Schack {\em et al.}~\cite{Schack} have shown that this formulation of quantum state estimation may be derived from the rules of quantum measurement theory, and may therefore be reformulated purely in terms of density matrices and POVM's. This formulation is as follows.

The initial state of the $N$ identically prepared systems may be written as 
\begin{equation}
  \rho^{(N)} =  \int P(\rho) \rho^{\otimes N} d\rho 
  \label{dfexp}
\end{equation}
where $\rho = |\psi\rangle\langle\psi |$, and $P(\rho)$ is the probability that the preparer choose to prepare all the systems in state $|\psi\rangle$. 
(In fact, due to the quantum de Finetti theorem (the details of which are 
given in ref.~\cite{Caves}), if we imagine the availability of an infinite 
sequence of these identically prepared systems, then the expansion given by 
Eq.(\ref{dfexp}) is unique for every value of $N$ (that is, for every subset 
of the infinite sequence). Thus, while we have started with a probability 
density over states, with certain assumptions even this can be replaced by 
an initial density matrix.)

Now, when an observer makes a measurement on one of the systems, the resulting state of the $N-1$ systems is given by~\cite{Schack}
\begin{equation}
   \rho^{(N-1)} =  \frac{1}{\cal N}\int P(\rho) \mbox{Tr}[E_k\rho] \rho^{\otimes (N-1)} d\rho 
\end{equation}
where $E_k$ is the effect associated with the particular measurement result, and ${\cal N}$ is the required normalization. Noting that $\mbox{Tr}[E_k\rho]$ is the conditional probability for the $k^{th}$ measurement result given that the density matrix for the measured system is $\rho$, we may write the final density matrix as \begin{equation}
  \rho^{(N-1)} =  \frac{1}{\cal N} \int P(\rho|k) \rho^{\otimes (N-1)} d\rho
\end{equation}
where 
\begin{equation}
  P(\rho|k) = \frac{1}{\cal N} P(k|\rho)P(\rho) .
\end{equation}
This last relation is simply Bayes' rule for updating a probability density $P$. Thus, if we like we can write the result of updating the observer's state of knowledge regarding $\rho$ by simply using Bayes rule on the initial probability density over $\rho$. Hence quantum state estimation can be written in exactly the same form as classical estimation of a classical variable.

To restate the problem of combining quantum states of knowledge in this setting, it is this: Given two observers, who have measured respectively $M$ and $L$ of the initial $N$ systems (for a total of $M+L$ measured systems), how do they combine their resulting states of knowledge about the remaining $N-M-L$ systems to obtain a total state of knowledge which includes correctly the information obtained by both. As in the classical problem, and the quantum problem considered in the previous section, to address this question we must decide what it means for the two observers to start with zero knowledge of the state to be inferred. So long as we restrict our attention to pure states, this question is easily answered. We simply choose the probability density for the state $\rho = |\psi\rangle\langle\psi |$ to be invariant under all unitary transformations~\cite{Jones}. To make the notation simple we choose the measure $d\rho$ to be invariant in this manner, and then the initial probability density $P = 1$. For convenience this invariant measure is given explicitly in Appendix~\ref{AppA}. However, the resulting initial density matrix for the $N$ systems is not proportional to the identity, which is the initial state which made sense in the previous analysis. This is easily seen by noting that if one were to start with the identity, a measurement on any given subsystem provides no information about the state of the remaining subsystems~\cite{Schack}, and therefore does not lead to the Bayesian update rule discussed above. As a result, this situation is not the same as that considered in the previous section.
  
It is now clear that if two observers have in their possession the probability densities describing their states of knowledge after their measurements, then they can easily combine these to form a total probability density, since this is exactly the same as the classical problem: the two observers simply multiply their densities together and normalize the result. However, what we are interested in here is whether two observers can combine their {\em density matrices} together to form a density matrix which is the result of two observers pooling their information, since in quantum mechanics it is the density matrix which captures the notion of a state of knowledge, and not the probability densities which appear in the formalism above. From what we know so far, it is fairly clear that this will not be the case. This is because the density matrices only consist of a small set of moments of the probability densities that generate them, and as a result we can choose many different probability densities consistent with a given density matrix. When we multiply these different probability densities together we can expect to get different combined states of knowledge, even if we fix the initial density matrices. We now provide an example to show that this is indeed the case. 

To do this we need to find two sets of measurement strategies which result in the same final two density matrices for the respective observers, but result in different combined states of knowledge. As the first set of measurements consider the situation where two observers, Alice and Bob, have initially three systems, and each measures one of them. This leaves one system about which they each have a state of knowledge. Alice's final state is
\begin{equation}
  \rho_{\mbox{\scriptsize A}} =  \frac{1}{{\cal N}_1} \int \mbox{Tr}[E_{k_1}^1\rho] \rho d\rho 
\end{equation}
where $E_{k_1}^1$ is the associated effect. While it is not necessary, for simplicity we will let Bob make the same measurement, and obtain the same measurement result as Alice, so that he has the same state of knowledge as her.
As the second case we consider the situation in which there are initially five systems, and Alice and Bob each measure two, leaving once again a single system. In this case Alice's final state is
\begin{equation}
  \rho'_{\mbox{\scriptsize A}} =  \frac{1}{{\cal N}_2} \int \mbox{Tr}[E_{k_2}^2\rho] \mbox{Tr}[E_{k_3}^3\rho] \rho d\rho 
\end{equation}
where $E_{k_1}^1$ and $E_{k_2}^2$ are the effects associated with the first and second measurements respectively, and again we will take Bobs final state to be the same. To make the example concrete, we take the systems to be two-state systems, and choose 
\begin{eqnarray}
  E_{k_1}^1 & = & A(\alpha) , \\
 E_{k_2}^2 & = & A(\beta) , \\
 E_{k_3}^3 & = & A(\gamma) ,  
\end{eqnarray}
where 
\begin{eqnarray}
 A(x) = \left( \begin{array}{cc} x & 0 \\ 0 & 1-x \end{array} \right) .
\end{eqnarray}
Parameterizing $\rho$ as 
\begin{eqnarray}
 \rho = \left( \begin{array}{cc} r & \sqrt{r(1-r)}e^{-i\phi} \\ \sqrt{r(1-r)}e^{i\phi} & 1-r \end{array} \right) ,
\end{eqnarray}
we have 
\begin{eqnarray}
  P(\rho|k_{1}) & = & \mbox{Tr}[E_{k_1}^1\rho] \nonumber \\ 
                & = & (2\alpha - 1) r + (1-\alpha) \\ 
  P(\rho|k_{2},k_{3}) & = & \mbox{Tr}[E_{k_2}^2\rho] 
                            \mbox{Tr}[E_{k_3}^3\rho] \nonumber \\ 
                      & = & (1-2\beta)(1-2\gamma) r^2 \nonumber \\
                      & - & [(1-2\beta)(1-2\gamma) + (1-\alpha-\beta)] r  \nonumber \\
                      & + & (1-\alpha)(1-\beta) 
\end{eqnarray}
and the final states are given by 
\begin{eqnarray}
 \rho_{\mbox{\scriptsize A}} & = &  \frac{1}{3}\left( \!\! {\small \begin{array}{cc} (\alpha+1) & 0 \\ 0 & \frac{1}{3}(2-\alpha) \end{array}} \!\! \right) , \\
 \rho'_{\mbox{\scriptsize A}} & = & \frac{1}{{\cal N}(\beta,\gamma)} \left( \!\! {\small \begin{array}{cc} (\beta\gamma +  \frac{1}{2})  & 0 \\ 0 & ((1-\beta)(1-\gamma) + \frac{1}{2}) \end{array}} \!\! \right) 
\end{eqnarray}
where ${\cal N}(\beta,\gamma) = 2(1-\beta)(1-\gamma) + (\beta + \gamma)$. In the first case, the state of knowledge of a third observer who has access to the results of the measurements of both observers (the combined state of knowledge), is
\begin{eqnarray}
 \sigma & = &   \frac{1}{{\cal N}(\alpha)} \int \mbox{Tr}[E_{k_1}^1\rho]^2 \rho d\rho \nonumber \\
          & = & \left( \begin{array}{cc} \frac{(\alpha^2 +  \frac{1}{2})}{2(1-\alpha)^2 + 2\alpha}  & 0 \\ 0 & \frac{((1-\alpha)^2 + \frac{1}{2})}{2(1-\alpha)^2 + 2\alpha} \end{array} \right) \\ 
\end{eqnarray}
and the second is 
\begin{eqnarray}
 \sigma' & = & \frac{1}{{\cal N}(\beta,\gamma)} \int (\mbox{Tr}[E_{k_2}^2\rho] \mbox{Tr}[E_{k_3}^3\rho])^2 \rho d\rho  
\end{eqnarray}
The explicit expression for $\sigma'$ is rather complex and we will not give it here, since it is sufficient for our purposes to evaluate the integral after putting in values for $\alpha$, $\beta$ and $\gamma$. 

We now wish to choose $\alpha, \beta$ and $\gamma$ so that $\rho_{\mbox{\scriptsize A}} = \rho^{'}_{\mbox{\scriptsize A}}$ and $\sigma \not= \sigma^{'}$. To satisfy the first condition, for a given $\alpha$ we must choose $\beta$ so that
\begin{equation}
  \beta = \frac{\frac{1}{3}(\gamma-2)(\alpha+1) + \frac{1}{2}}{\frac{1}{3}(2\gamma-1)(\alpha + 1) - \gamma}
\end{equation}
which leaves $\gamma$ as a free parameter. Choosing $\gamma$ to satisfy the second condition is not hard. For example, if we take $\alpha = 1/2$ and $\gamma=1/4$ which gives $\beta = 3/4$, we have 
\begin{eqnarray}
 \rho_{\mbox{\scriptsize A}} & = & \rho'_{\mbox{\scriptsize A}} = \frac{1}{2}I ,
\end{eqnarray}
where as 
\begin{eqnarray}
 \sigma & = & \frac{1}{2}I
\end{eqnarray}   
and 
\begin{eqnarray}
 \sigma' & = & \frac{1}{406}\left( \begin{array}{cc} 299 & 0 \\ 0 & 107 \end{array} \right)  
\end{eqnarray}
Thus even for quantum state-estimation, the state resulting from the combined information of two observers is not well defined by their respective density matrices alone.

\section{Conclusion}
While combining independently obtained states of knowledge is simple in classical statistical inference, the problem is significantly more complex in quantum mechanics. In particular, the state of knowledge possessed by a third observer who has access to the combined information of two independent observers can be any state who's support lies in the intersection of the supports of the states of the two observers. Thus only in the special case in which this intersection is one dimensional is the state resulting from full information well defined by the two observers' states alone. Nevertheless, we stress that it may well still be possible to obtain a unique, well-motivated rule for combining density matrices. One way to obtain such a rule might be to average over all the possible measurement histories consistent with the final two states of knowledge, weighted by the relevant conditional probability for each history. To do this however, a well motivated measure over measurement histories would need to be found.

\section*{Acknowledgments}
We would like to thank Salman Habib for helpful discussions and a thorough reading of the manuscript. This research is supported by the Department of Energy, under contract W-7405-ENG-36.

\appendix
\section{Invariant measure}
\label{AppA}
The measure over $d$-dimensional pure states, $|\psi\rangle$, which is invariant under all unitary transformations, is~\cite{Jones} 
\begin{equation}
   2\frac{\pi^n}{(n-1)!}dx_1 \;\ldots\; dx_d \; dy_1 \;\ldots\; dy_d
   \label{meas1}
\end{equation}
where
\begin{equation}
   |\psi\rangle = \sum_{k=1}^d (x_k + iy_k) |k\rangle, 
\end{equation}
in which the set $\{|k\rangle\}$ is a basis for the system, and the integration is performed over the surface of a $2d$-dimensional hypersphere given by $\sum_{k=1}^d x_k^2+y_k^2 = 1$. For our purposes it is more useful to have this in terms of probabilities, $P_k$, and phase angles $\theta_k$, where 
\begin{eqnarray}
   P_k & = & x_k^2 + y_k^2 \\   \theta_k & = & arg(x_k + iy_k) .
\end{eqnarray}
To change to these variables we first write the integral as a volume integral using the identity~\cite{GradRys}
\begin{eqnarray}
 & & \int_S f(\{x_i,y_i\}) \; dx_1\ldots  dx_d \; dy_1\ldots  dy_d = \\ \nonumber 
& & \int_{R_1}\!\!\!\! (f(y_d=\kappa) +f(y_d=-\kappa))\frac{1}{\kappa}\prod_{i=1}^{d}dx_i\prod_{i=1}^{d-1}dy_i, \label{ints}
\end{eqnarray}
where
\begin{equation}
 \kappa = \left( 1\!\! - \!\! \sum_{i=1}^{d} x_i^2 - \!\! \sum_{i=1}^{d-1}y_i^2\right)^{-1/2}
\end{equation}
and $R_1$ is given by $\{\sum_{i=1}^{d} x_i^2+\sum_{i=1}^{d-1}y_i^2 \leq 1\}$. The first term in the integral corresponds to the top half of the sphere,and the second term to the bottom half. In what follows we will only perform the change of variables for the top half, since the bottom half gives naturally the same result for the measure. One must merely remember to use $y_d=-\kappa$ as the argument for $f$ when integrating over the bottom half of the sphere. As the prior probability is unity, for the case we are interested in, $f=1$ and so is independent of $y_d$. We now make a change of variables to $\{r_i,\theta_i\}$ where\begin{equation}
  x_i = r_i\cos(\theta_i) \; , \;\; y_i = r_i\sin(\theta_i)
\end{equation}
for $i$ in the range $1\ldots d-1$. This gives 
\begin{equation}
\!\!\int_0^{2\pi}\!\!\!\! \!\! \cdot \!\! \cdot \!\! \cdot \!\!\int_0^{2\pi}\!\!\!\!\int_{R_2}\!\!\!\! f \frac{1}{\kappa}\; dx_d \prod_{i=1}^{d-1}r_i\;dr_i \prod_{i=1}^{d-1} d\theta_i ,
\end{equation}
where $\kappa$ may be written as 
\begin{equation} 
\kappa = \left( 1 \!\! - \!\! \sum_{i=1}^{d-1} r_i^2- x_d^2 \right)^{-1/2}
\end{equation}
and $R_2$ is given by $\sum_{i=1}^{d-1}r_i^2+x_n^2\leq 1$. We next make a change of variable from $x_n$ to $\theta_n$ where
\begin{equation}
  x_n = \rho \sin{\theta_n},\end{equation}and\begin{equation}  \rho^2 = 1-\sum_{i=1}^{d-1}r_i^2 .
\end{equation}
Noting that 
\begin{equation}
  \kappa = \rho |\cos{\theta_n}|,
\end{equation}
the integral becomes
\begin{equation}
\int_0^{2\pi}\!\!\!\! \!\! \cdot \!\! \cdot \!\! \cdot \!\!\int_0^{2\pi} \int_{-\pi/2}^{\pi/2} \left\{ \int_{R_3}\!\!\!\! f \,\prod_{i=1}^{d-1}r_i\; dr_i \right\} d\theta_d \prod_{i=1}^{d-1} d\theta_i,
\end{equation}
where $R_3$ is given by $\sum_{i=1}^{d-1}r_i^2 \leq 1$. Finally, changing variables from the $r_i$ to $P_i=r_i^2$, we have
\begin{equation}
\int_0^{2\pi}\!\!\!\! \!\! \cdot \!\! \cdot \!\! \cdot \!\!\int_0^{2\pi} \int_{-\pi/2}^{\pi/2} \left\{\frac{1}{2^{d-1}}\int_{R_3}\!\!\!\! f \,\prod_{i=1}^{d-1}dP_i \right\} d\theta_d \prod_{i=1}^{d-1} d\theta_i ,
\end{equation}
where in terms of the probabilities $P_i$, the region $R_3$ is given by $\sum_{i=1}^{d-1}P_i \leq 1$.

\end{document}